\documentclass[%
onecolumn,
 amsmath,amssymb,
 aps,
]{revtex4-2}

\usepackage{xcolor}

\usepackage{graphicx}
\usepackage{dcolumn}
\usepackage{bm}
\usepackage{comment}

\begin{document}

\preprint{APS/123-QED}

\title{Critical active dynamics is captured by a colored-noise driven field theory}  

\author{Claudio Maggi}
\email{claudio.maggi@cnr.it}
\affiliation{NANOTEC-CNR, Institute of Nanotechnology, Soft and Living Matter Laboratory, Roma, Italy.}
\affiliation{Dipartimento di Fisica, Sapienza Universit\`a di Roma, Piazzale A. Moro 2, I-00185, Rome, Italy.}

\author{Nicoletta Gnan}
\affiliation{ISC-CNR, Institute for Complex Systems, Piazzale A. Moro 2, I-00185 Rome, Italy.}
\affiliation{Dipartimento di Fisica, Sapienza Universit\`a di Roma, Piazzale A. Moro 2, I-00185, Rome, Italy.}

\author{Matteo Paoluzzi}
\affiliation{Departament de Física de la Mat\`eria Condensada, Universitat de Barcelona, C. Martí Franqu\`es 1, 08028 Barcelona, Spain.}

\author{Emanuela Zaccarelli}
\affiliation{ISC-CNR, Institute for Complex Systems, Piazzale A. Moro 2, I-00185 Rome, Italy.and\\ Dipartimento di Fisica, Sapienza Universit\`a di Roma, Piazzale A. Moro 2, I-00185, Rome, Italy.}

\author{Andrea Crisanti}
\affiliation{Dipartimento di Fisica, Sapienza Universit\`a di Roma, Piazzale A. Moro 2, I-00185, Rome, Italy.}

\begin{abstract}
We numerically investigate the correlation function, the response and the breakdown of the Fluctuation-Dissipation Theorem (FDT) in active particles close to the motility-induced critical point. By performing extensive computer simulations we find a strong FDT violation in the high-wavevector and high-frequency regime where the frequency-resolved response function has a much larger amplitude than the spectrum of spontaneous fluctuations. Conversely, upon monitoring the dynamics at large spatiotemporal scales, the response and the correlator progressively become closer, restoring the FDT and implying that the critical dynamics at large scales is in effective equilibrium. Moreover we show that this critical slowing-down is compatible with the Ising universality class with conserved order parameter.
Building on these numerical results, we develop a novel field-theoretical description employing a space-time correlated noise field and we show that the theory qualitatively captures the numerical results already at the Gaussian level. We further perform a one-loop dynamic renormalization group analysis to demonstrate that the correlated noise does not change the critical exponents with respect to the equilibrium theory. Finally we show that our theoretical model gives a frequency-resolved effective temperature which is in good agreement with the simulation data of the microscopic model. Our results demonstrate that a correlated noise field is a fundamental ingredient to capture the non-equilibrium features of critical active matter at the coarse-grained level.
\end{abstract}

\maketitle


\section*{Introduction}

Active matter may display striking non-equilibrium phenomena such as the unidirectional propulsion of ratchet motors driven by active particles~\cite{di2010bacterial,maggi2016self} or the spontaneous accumulation of self-propelled bacteria or colloids interacting with asymmetric obstacles~\cite{galajda2007wall}.
However, there are situations where the non-equilibrium features of an active system are not immediately evident as, for instance, when self-propelled particles  exhibit a collective motion or self-organization on large scales~\cite{Marchetti13} similar to what observed in equilibrium systems. The archetypal example of this kind of behavior is the Motility-Induced Phase Separation (MIPS)~\cite{cates2015motility}. This phenomenon can be microscopically understood in terms of active particles that move slower in denser regions, thus triggering an effective attraction that brings the system close to a spinodal decomposition~\cite{PhysRevE.48.2553, Tailleur08, PRLMarchetti2012, stenhammar2013continuum, speck2014effective, GONNELLA2015,  siebert2017phase, PhysRevLett.122_2019, mandal2019motility}. Although the MIPS can be observed even in active particles without attractive interaction forces, it shares many similarities with the gas-liquid coexistence in equilibrium systems. Such analogies can be captured through effective equilibrium approaches which allow to reduce the non-equilibrium fluctuating forces to an effective interaction potential~\cite{cates2015motility,Farange15,maggi2015multidimensional,Marconi3,Marconi17,PhysRevResearch.2.023207}. The mapping to the gas-liquid phase transition suggests that MIPS ends in a critical point. Although previous numerical results have suggested that, in some active systems, such a critical point displays non-Ising exponents~\cite{siebert2018critical}, recent studies have shown that some other active on-lattice and off-lattice models have a critical point belonging to the Ising universality class~\cite{PhysRevLett.123.068002,maggi2021universality}.
As a consequence in the latter case, equilibrium-like 
$\varphi^4$ field theories~\cite{ZinnJustin} should provide the correct asymptotic description of MIPS around its critical point, at least at sufficiently large spatial dimensions. However, it has been pointed out that additional terms, which break the time reversal symmetry, should be included in the standard $\varphi^4$ theory to capture the non-equilibrium and non-universal features near the MIPS critical point. Although these terms turn out to be irrelevant from the point of view of Renormalization Group (RG) transformations, they might yield, for example, a non-zero entropy production rate~\cite{PhysRevLett.124.240604}. 
In this context it is thus crucial to understand if and how the critical dynamics of an active system becomes effectively identical to the one of an \textit{equilibrium} critical system.
 
One of the most direct and natural way to unveil the non-equilibrium nature of a system is to look at the response and the correlation function of the observable of interest. 
These dynamical functions are related in equilibrium by the Fluctuation Dissipation Theorem (FDT). Its violation can be thus used to quantify how far the system is from equilibrium at various spatio-temporal scales.
This approach has been widely employed to study the properties of several off-equilibrium systems such as glasses, gels and granulars~\cite{BarratFDT, Bellon_2001, puglisi2002fluctuation, crisanti2003violation, kurchan2005, potiguar2006effective, gnanFDT2010,  maggi2010generalized, GnanFDT2011, Cugliandolo_2011}. Simultaneous measurements of response and correlation functions have also been used to reveal non-equilibrium fluctuations in active particle simulations~\cite{burkholder2019fluctuation,Fodor16,loi2008effective,dal2019linear,caprini2021fluctuation,szamel2017evaluating} and in active experimental systems, such as living red-blood cell membranes~\cite{turlier2016equilibrium} and suspensions of swimming bacteria probed by passive tracers ~\cite{chen2007fluctuations,maggi2017memory}.

The aim of the present work is to collect and use the information from the critical correlation and the response of the order parameter field, in a data-driven approach, to build a field-theoretical model that is able to faithfully reproduce the non-universal features of active particles close to the MIPS critical point. To carry out this program, we perform numerical investigations based on an efficient field-free method allowing us to compute correlation and response functions at criticality probing different length and time scales \cite{szamel2017evaluating}. 
We find that the FDT is strongly violated at high frequencies and large wave-vectors. In this regime we find that the frequency-resolved correlator is much lower in amplitude than the corresponding response. Upon lowering the frequency and the wave-vector, we find that the response and correlator tend to coincide, thus satisfying FDT and validating the effective equilibrium picture.

To rationalize this numerical evidence, we put forward a colored-noise driven dynamical field theory, that is able to explain the coarse-grained behavior of the active system. In particular we show that our results can be qualitatively explained by a scalar theory with conserved order parameter (Model B) and driven by a random field that is correlated both in space and in time. Interestingly, we find that already at the Gaussian level (i.e. ignoring the effect of fluctuations), our theory predicts a scale-dependent violation of FDT that is restored on large length- and time-scales in full agreement with the simulations. By further investigating this field-theoretical framework, we observe that the additional parameters describing the noise field -- its spatial correlation length and correlation time  -- are irrelevant under RG transformation and the system belongs to the Ising universality class. Additionally, by taking into account the non-linear coupling, we show that the critical exponents do not change up to first-order in Wilson's RG, suggesting that the 
theory is consistent with the expected asymptotic equilibrium-like criticality, also close to the upper critical dimension. 
With respect to the active field theories considered so far, our model captures the non-equilibrium features of critical active particles already by considering only linear terms and without the addition of non-integrable (and non-linear) contributions to the field dynamics~\cite{nardini2017entropy,PhysRevLett.124.240604,caballero2018bulk}. The overall picture, stemming from simulations and theory, unveils that a colored noise-field is a fundamental ingredient in the field-theoretical framework  that is needed to capture the non-universal features of near-critical active fluids.

\section*{Results}

\subsection*{Microscopic model}

\begin{figure}[!t]
\includegraphics[width=0.375\columnwidth]{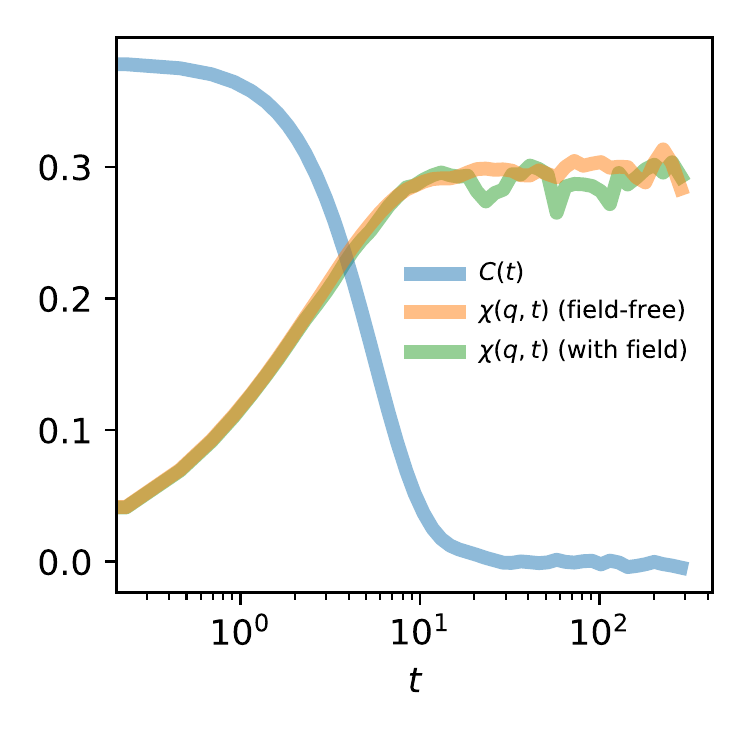}
\caption{\label{fig:fieldfree} \textbf{Perturbed and spontaneous dynamics of active particles.}
The field-free method allows to extract the correlation function of the spontaneous fluctuations (blue curve) and the integrated response function (orange curve). The latter coincides with the response obtained by directly applying the field (green curve).}
\end{figure}

We consider a system composed by $N$ self-propelled active Ornstein–Uhlenbeck particles~\cite{maggi2015multidimensional,SFB15} (AOUPs) in $d=2$. This model has been intensively studied for its remarkable analytic properties~\cite{PhysRevLett.117.038103,dal2019linear,bonilla2019active,wittmann2017effective,marconi2016effective,Paoluzzi16} and, as discussed in the following, it is the ideal model for characterizing the perturbed and spontaneous critical dynamics. The equations of motion of AOUPs read

\begin{eqnarray}
\label{micro1}
\dot{\mathbf{r}}_i &=& \mu \, (  \boldsymbol{\psi}_i +  \, \mathbf{F}_i ) \\ 
\label{micro2} 
\tau \, \dot{\boldsymbol{\psi}}_i &=& -\boldsymbol{\psi}_i + \boldsymbol{\xi}_i
\end{eqnarray}

\noindent where $\mathbf{r}_i$ indicates the $i$-th particle's position and $\boldsymbol{\psi}_i$ is the self-propelling force. $\mathbf{F}_i = \sum_{j\neq i} \mathbf{f}_{ij}$ represents the force acting on the particle generated by two body interactions, i.e., $\mathbf{f}_{ij} = -\nabla_{\mathbf{r}_i} \phi(r_{ij})$, with $r_{ij} = | \mathbf{r}_i - \mathbf{r}_j| $. The two-body potential is modeled as a steep inverse power-law $\phi(r)=(r/\sigma)^{-12}/12$ with a cut-off at $r=2.5\,\sigma$. Here $\sigma$ represents the diameter of the particle and is set to 1.
In Eq.~(\ref{micro2}) $\tau$ is the persistence time of the active force and $\boldsymbol{\xi}$ is a standard white noise source, i.e. $\langle \xi_{i}^\alpha(t) \rangle=0$ and $\langle \xi_{i}^\alpha(t) \xi_{j}^\beta(s) \rangle = 2 (D/\mu^2)\, \delta_{ij} \delta(t-s)  $, where the greek indices indicate the Cartesian components. 
Here $D$ is the diffusivity of the non-interacting particles and $\mu$ is the particles mobility (set to 1 in simulations). 
Note that, by taking the limit $\tau \rightarrow 0$ at constant $D$, AOUPs reduce to passive Brownian particles in equilibrium at temperature $T=D/\mu$ ($k_B=1$ units are used throughout the present work).

We have recently located the MIPS critical point for this  model and demonstrated that the resulting \textit{static} critical exponents are in agreement with the Ising universality class~\cite{maggi2021universality}.
In the present work we focus our study on the dynamics of the near-critical state-point with density $\rho=0.95$ 
and active-force parameters ${\tau=D=16.5}$ and by employing various system sizes. Particles are enclosed in a rectangular box of sides ${(L_x,L_y=L_x/3)}$ and periodic boundary conditions apply, as in Ref.~\cite{maggi2021universality}.
Being the density the order parameter field of our critical system, we study the real part of its Fourier transform at time~$t$, i.e.  ${\rho_\mathbf{q}(t) = \sum_i \cos(\mathbf{q}\cdot \mathbf{r}_i(t))}$, where the sum runs from $i=1$ to $i=N$ and ${\mathbf{q}=2 \pi (n_x/L_x,n_y/L_y)}$ represents the wave-vector (here ${n_{x,y}\in \mathbb{Z}}$). The quantity which statistically characterizes the spontaneous fluctuations of the field is the
auto-correlation function ${C(q,t)=2 N^{-1}\langle \rho_{\mathbf{q}}(t+s) \, \rho_{\mathbf{-q}}(s) \rangle}$
where brackets indicate the average over configurations.
The associated linear response function is ${\chi(q,t) = N^{-1}[\partial_h \langle \rho_\mathbf{q}(t)\rangle]_{ h \rightarrow 0}}$ where $h$ is the amplitude of an external force field $\mathbf{f}_i^\mathrm{ext}$ which perturbs the $i$-th particle dynamics of Eq.~(\ref{micro1}) as
$
\dot{\mathbf{r}}_i=   \mu \, ( \boldsymbol{\psi}_i + \, \mathbf{F}_i +  \mathbf{f}_i^\mathrm{ext})
$, and has the form
${
\mathbf{f}_i^\mathrm{ext}(t) = -2\,h \, \Theta(t) \, \mathbf{q} \, \sin(\mathbf{q}\cdot\mathbf{r}_i(t))
}
$
with $\Theta(t)$ being the Heaviside step function.

The definition of $\chi(q,t)$ immediately poses two main difficulties. The first is that computing $\chi(q,t)$ at different $q$-values requires a different simulation for each $q$.
Secondly, to ensure that the field amplitude is small enough to avoid non-linear effects, one should repeat the simulations for various values of $h$. Fortunately, applying the field is not required for measuring $\chi(q,t)$ in AOUPs. Indeed, among the ``family'' of active models with exponentially correlated noise~\cite{koumakis2014directed}, AOUPs have the unique feature that the exact linear response of any dynamic variable can be computed in \textit{unperturbed} simulations as demonstrated in Ref.s~\cite{szamel2017evaluating,caprini2021fluctuation}. In the present work we employ the method developed by Szamel~\cite{szamel2017evaluating}, which generalizes the Malliavin weights method~\cite{bell2006malliavin} to systems driven by persistent Gaussian noise and efficiently combines it with parallel (GPU-based) simulations. The key idea of the method is to derive the appropriate fluctuating variables, encoding the linear response, by taking the limit of vanishing external field \textit{before} taking the statistical average.
Details on the method implementation can be found in Appendix ~\ref{MalliApp}.
Fig.~\ref{fig:fieldfree} shows the functions of interest $C(q,t)$ and $\chi(q,t)$ at $\mathbf{q}=2\pi(6/L_x,2/L_y)$ for a system of $N=3750$ at ${(\rho = 0.95,\tau=D=16.5)}$. The response function $\chi(q,t)$ is evaluated both with a small external field ($h=0.1$) and with the field-free method,  showing a good agreement between the two and validating the implementation of the latter technique.


\subsection*{Correlation and response functions}

\begin{figure*}[!ht]
\includegraphics[width=1.\textwidth]{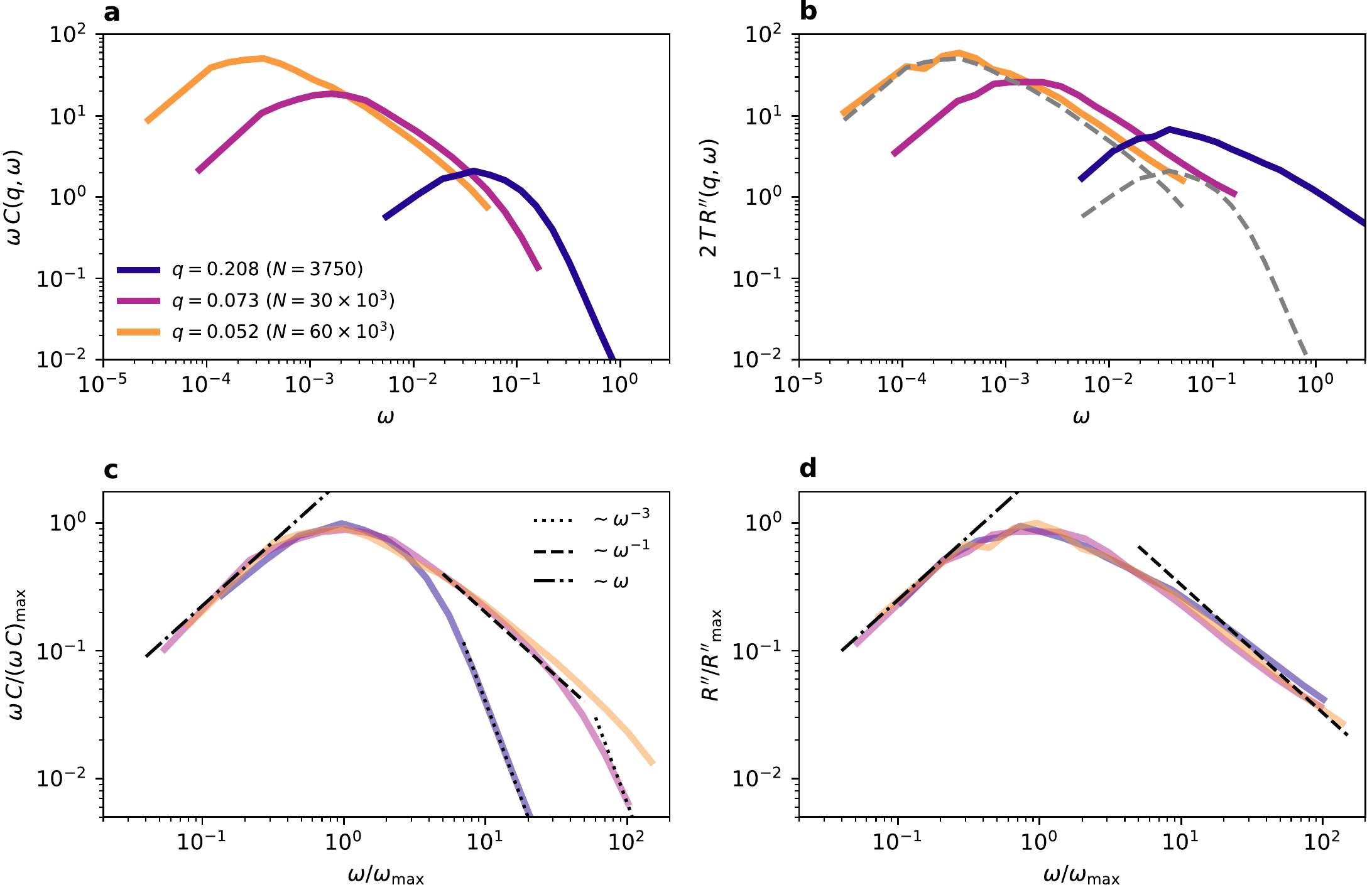}
\caption{\label{fig:wide1} \textbf{Correlation and response in the critical active system.} (\textbf{a}) Fluctuation spectrum as a function of frequency $\omega$ at various $q$-s (colored curves). Simulating larger systems gives access to lower wave-vectors (see legend). (\textbf{b}) Frequency-resolved response function evaluated by the field-free method at the same $q$-s as in (a) (colored curves, same legend as in (a)). The correlation spectra for the highest and lowest wave-vectors in (a) are reproduced here as dashed lines to show that the deviation between response and correlation is more pronounced at high $q$.
(\textbf{c}) Scaled fluctuation spectra from (a) revealing an evident deviation from scaling in the high-$q$/high-$\omega$ regime. The asymptotic behaviors of these functions are highlighted by power-law fits (straight lines, see legend).
(\textbf{d}) Scaled response functions from (b). Differently from (c), a good data-collapse is observed for the response function. The asymptotic behaviors are evidenced by power law fits (straight lines, same legend as in (c)). 
}
\end{figure*}

We report here the results for correlation and response of the critical active system at different spatio-temporal scales. 
We focus on the frequency-resolved correlation and response functions, which we indicate as $C(q,\omega)$ and $R(q,\omega)$ respectively, that are obtained as the time-Fourier transforms ${C(q,\omega)=\int dt \, e^{i\omega t} \, C(q,t)}$ and ${R(q,\omega)= i \, \omega \, \int dt \, e^{i\omega t} \, \chi(q,t)}$. Note that using the factor $i\,\omega$ yields the frequency-resolved \textit{impulsive} response function which, in equilibrium, is related to $C(q,\omega)$ by the FDT: 
$\omega \, C(q,\omega) = 2 \, T R''(q,\omega)$, where the double prime indicates the imaginary part.
Fig.~\ref{fig:wide1}(a)-(b) reports the frequency-resolved correlation and response functions at the near-critical state-point ${(\rho=0.95,\tau=D=16.5)}$.
To access a wide range of $q$-values we employ different system sizes ranging from $N=3750$ to $N=60 \!\times\! 10^3$, rather than simulating one single large system, in order to speed-up the simulations at high $q$-values. 
Fig.~\ref{fig:wide1}(a) shows the evolution of  $\omega\,C(q,\omega)$ for different $q$-values. We observe that the correlator grows in amplitude and its peak shifts to lower frequencies upon lowering $q$, signalling a considerable slowing-down at large length-scales.

Fig.~\ref{fig:wide1}(b) shows  that the response $2\,T\,R''(q,\omega)$ has a similar qualitative behavior. However, by reporting $\omega\,C(q,\omega)$ on the same plot (gray dashed lines Fig.~\ref{fig:wide1}(b)), it is immediately evident that the response has a much higher amplitude than the correlator at large $q$-values, implying a dramatic violation of the FDT at these length-scales. Moreover, the high-$q$ correlator decays faster than the response at high $\omega$, further suggesting that the FDT-violation is also frequency-dependent. Differently, we observe that at low $q$-values
$\omega\,C(q,\omega)$ and $2\,T\,R''(q,\omega)$ become much closer and almost perfectly coincide for small frequencies. 
To appreciate the different frequency-dependence between correlation and response, we plot these functions scaled by their respective maximum in Fig.~\ref{fig:wide1}(c) and (d).
Interestingly we find that, while the shape of $\omega\,C(q,\omega)$ significantly changes upon changing $q$, $R''(q,\omega)$ does not vary and a good collapse is found for the scaled response at all $q$-s. 
More specifically, in the low-$\omega$ regime, both correlation and response show a good collapse and they both grow as $\omega$ as highlighted by the dashed-dotted power law fits in Fig.~\ref{fig:wide1}(c) and (d). Differently, in the high-$\omega$ region, the response decays slightly slower than $\omega^{-1}$ at all $q$-s, while the correlator behavior shows a progressive change from a single $\omega^{-3}$ decay at high $q$ (purple curve in Fig.~\ref{fig:wide1}(c)) to a two-step decay at intermediate $q$ (magenta curve), first showing a $\omega^{-1}$ dependence followed by a crossover to a faster decay. Finally, at very low $q$ (orange curve), the decay is almost fully captured by a slow (approximately $\omega^{-1}$) power-law.

In addition, the peak positions of $\omega \, C(q,\omega)$ and $R''(q,\omega)$, which can be identified as the system relaxation frequency, also show some differences. We denote the frequency where the peak is found as $\omega_\mathrm{max}$ and we plot it as a function of $q$ for both functions in Fig.~\ref{fig:wide2}(a). We observe that the correlator has an $\omega_\mathrm{max}$ lower than the one of the response in the high-$q$ regime, implying that the response relaxes faster than the correlator at these length scales.
However, at small $q$-s the two relaxation frequencies almost perfectly coincide, nicely following a power-law decay as $\omega_\mathrm{max}~\sim q^z$.
A direct power law fit of the low $q$-values (orange points in Fig.~\ref{fig:wide2}(a)) yields $z=3.78(0.13)$ for the response and $z=3.80(0.14)$ for the correlator (the fit error is reported in brackets) in good agreement with the critical exponent $z=3.75$ of the equilibrium Ising model with conserved magnetization in $d=2$~\cite{hohenberg1977theory}.
The deviation from the power-law behavior observed at large $q$-values signals that we are probing the microscopic relaxation and hence we are far from the scaling regime.
To better understand this deviation, we have also simulated the critical equilibrium triangular lattice gas~\cite{maggi2021universality}, shown in the inset of Fig.~\ref{fig:wide2}(a)), which displays a qualitative similar deviation at large $q$-s. This suggests that, independently on the fact the the system is in equilibrium or not, at large enough $q$-values one is probing the dynamics at short scales where the microsocpic details play an important role and the relaxation frequency does not follow a scaling law.

To conclude this paragraph it is worth noting that, by using the active field theoretical framework proposed in~\cite{nardini2017entropy,PhysRevLett.124.240604,caballero2018bulk}, one could in principle evaluate the entropy production rate directly from the numerical correlation and response exploiting the Harada-Sasa (HS) relation~\cite{harada2005equality}.
However, our numerical results show that, while $\omega \, C(q,\omega)$ decays relatively fast ($\omega \, C(q,\omega) \sim\omega^{-3}$) at high frequencies, the response $R''(q,\omega)$ decays much slower ($R''(q,\omega) \sim \omega^{-1}$). This implies that the entropy production rate $\mathcal{S}$ \textit{cannot} be computed using the HS relation, i.e. as done in Ref.~\cite{nardini2017entropy} by employing the formula
${\mathcal{S} \propto \int dq \, q^{d-1} \int_{-\infty}^{\infty} d\omega \, q^{-2}\omega\,[\omega \, C(q,\omega)-2 \,T\, R''(q,\omega)]}$, since this quantity \textit{{diverges}} upon performing the frequency integral. 
Interestingly, a similar divergence is obtained when studying one single harmonic oscillator driven by a colored (Ornstein-Uhlenbeck) noise, in which one also finds that ${R''(\omega)\sim \omega^{-1}}$ and ${\omega\, C(\omega)\sim \omega^{-3}}$.
These results suggest that (\textit{i}) any field theory, including only white noise, is not sufficient to capture the behavior observed in numerical simulations and (\textit{ii}) that our numerical results could be rationalized in terms of colored noise. We will return to this point in more detail at the end of the next paragraph after showing further evidenced in favour of a theory characterized by colored noise.

\subsection*{Effective temperature}

From the observations presented above it emerges that the FDT violation depends on both $q$ and $\omega$.
It is therefore desirable to quantify the FDT violation by a single function, which we identify as the (normalized) frequency-dependent ``effective temperature'':

\begin{equation}\label{eq:teffdef}
T_{\rm eff}(q,\omega)=
\frac{ \omega \, C(q,\omega)}{2\, T\,R''(q,\omega)} \; .
\end{equation}

\noindent The normalization ensures that, when $T_{\rm eff}(q,\omega)=1$, the system can be considered as being in effective equilibrium at the bath temperature $T$.
We emphasize that here we do not assign any deeper meaning to the $T_{\rm eff}$ than a convenient ``violation factor''. While it has been shown that, in some glassy systems, the separation of timescales of aging processes allows one to interpret $T_{\rm eff}$ as a ``thermodynamic'' temperature~\cite{sciortino2001extension,crisanti2003violation}, here we rather use Eq.~(\ref{eq:teffdef}) only as a suitable measurement of the FDT violation.

\begin{figure*}[!t]
\includegraphics[width=\textwidth]{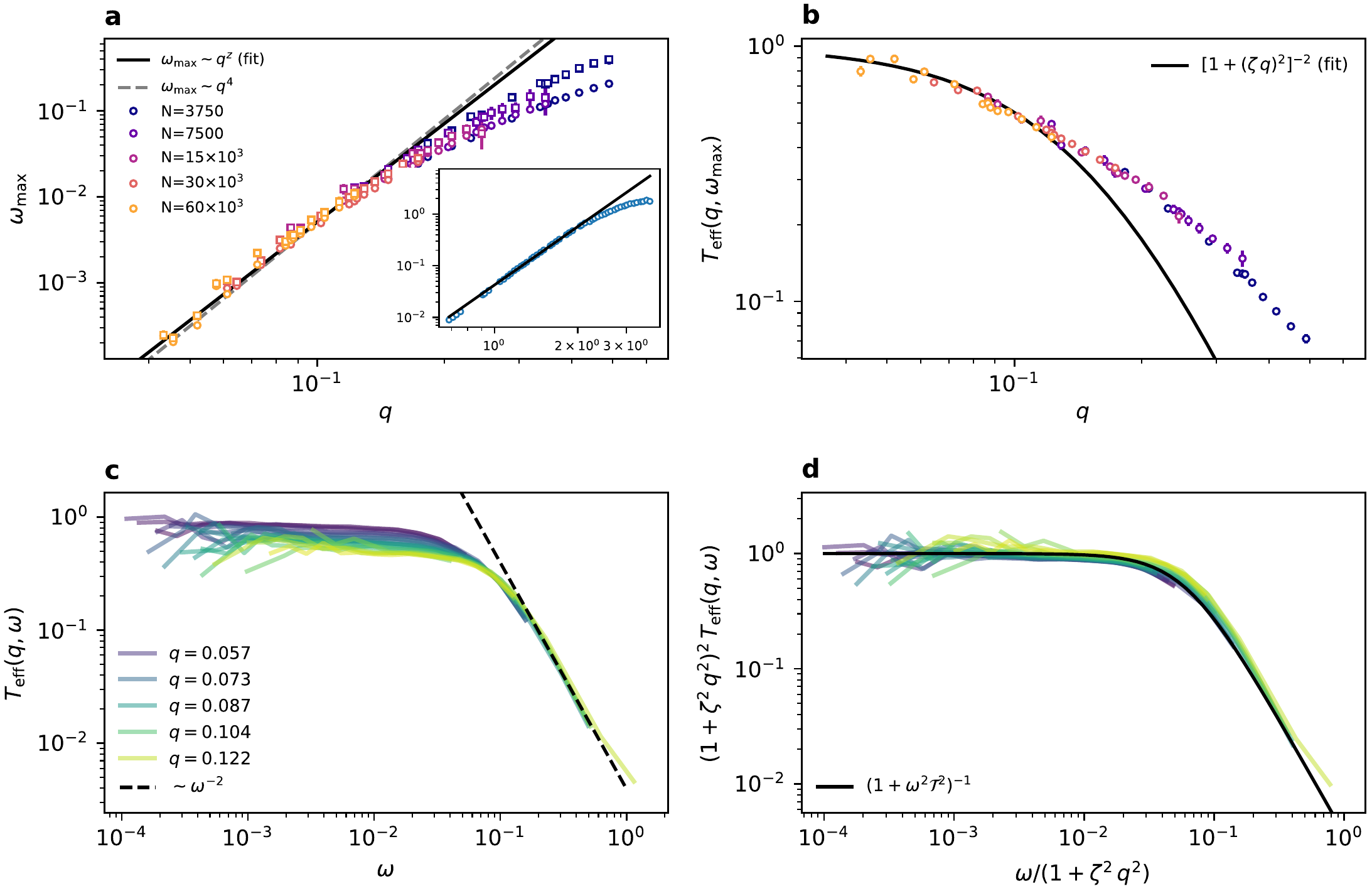}
\caption{\label{fig:wide2} \textbf{Relaxation frequency and effective temperature of the critical active system.} (\textbf{a})
Peak frequency $\omega_\mathrm{max}$ of the correlator (circles) and of the response (squares)
as a function of $q$. Different colors indicate different systems sizes (see legend). The straight full line is a power-law fit of the low-$q$ correlator peak-frequencies (i.e. the yellow points for which $N=60\times 10^3$), yielding a critical exponent $z$ very close to the Ising one ($z=3.75$), which is lower than the mean-field value (the best fit with $z=4$ is indicated by the dashed line).  
The inset (same axes as the main panel) shows the peak frequency for the equilibrium triangular lattice gas. Deviations from the scaling $\omega_\mathrm{max}\sim q^4$ are also found at high $q$-values. 
(\textbf{b})
Effective temperature evaluated at the correlator peak-frequency $T_{\rm eff}(q,\omega)_\mathrm{max}$ as a function of $q$ (different colors indicate different system sizes with the same legend as in (a)). The black curve is a fit of the low-$q$ data by means of the expression found in the one-loop colored-noise driven field theory (Eq.~(\ref{eq:Teff})).
(\textbf{c})
Frequency resolved effective temperature $T_{\rm eff}(q,\omega)$ in the low-$q$ regime (colored curves), different colors indicate different $q$-values (see legend). The dashed line represents the power-law $\omega^{-2}$ predicted by the theory for asymptotic behavior of $T_\mathrm{eff}$.
(\textbf{d})
data in (c) are scaled according to the theory. Collapsed data are well fitted by the Lorentzian predicted by the theory (full line).
}
\end{figure*}

To first understand the $q$-dependence of $T_{\rm eff}$, we show in Fig.~\ref{fig:wide2}(b) the quantity $T_{\rm eff}(q,\omega_{\rm max})$ that is the $T_{\rm eff}$ evaluated at the characteristic frequency $\omega_{\rm max}$ of the correlator. We find that the large-$q$ regime is dominated by a strong violation of the FDT with a $T_{\rm eff}(q,\omega_{\rm max})< 1$ which can be interpreted as if,  on small length scales, the systems appears to fluctuate with a temperature lower than the bath temperature. Upon lowering $q$, however, $T_{\rm eff}(q,\omega_{\rm max})$ asymptotically tends to  unity suggesting that, over large scales, the system relaxes as if it were in equilibrium at temperature $T$. 
Moreover, to understand the $\omega$-dependence of $T_\mathrm{eff}$, we plot $T_\mathrm{eff}(q,\omega)$ as a function of $\omega$ in Fig.~\ref{fig:wide2}(c). We note that, even at low $q$, where the relaxation frequencies of response and correlator coincide, $T_{\rm eff}(q,\omega)$ still displays clear violation of the FDT at high $\omega$,
where the decay of $T_{\rm eff}(q,\omega)$ is well captured by a power law $ \sim \omega^{-2}$. This result is consistent with Eq.~(\ref{eq:teffdef}) considering that, at high frequencies, we have found the approximate decays  $\omega \, C(q,\omega) \sim \omega^{-3}$ and $R''(q,\omega) \sim \omega^{-1}$ (see Fig.~\ref{fig:wide1}(c) and (d)). Differently, for $\omega$ smaller than $\omega_\mathrm{max}$, $T_{\rm eff}(q,\omega)$ remains close to unity and progressively approaches one as $q$ decreases.

To summarize our numerical findings, we found that the spontaneous fluctuations (measured by the correlator) are \textit{weaker} than those induced by the external field (measured by the response) at short time- and length-scales. These  quantities allow us to investigate the microscopic model on a coarse-grained level in terms of the density field having its own non-equilibrium fluctuating dynamics.
The scenario found for these density relaxation functions can thus be modeled by considering a coarse-grained noise-field which does not ``excite'' enough the density modes at high $q$ and $\omega$ and therefore is ``colored''. Differently the external field, inducing the response, can vary on an arbitrarily high frequency and wavelength equally exciting all available modes. As discussed in the following paragraph, this picture can be made rigorous by means of a field-theoretic approach and describes qualitatively well the simulation data.

\subsection*{Colored-noise driven field theory}

\begin{figure*}[!ht]
\includegraphics[width=1.\textwidth]{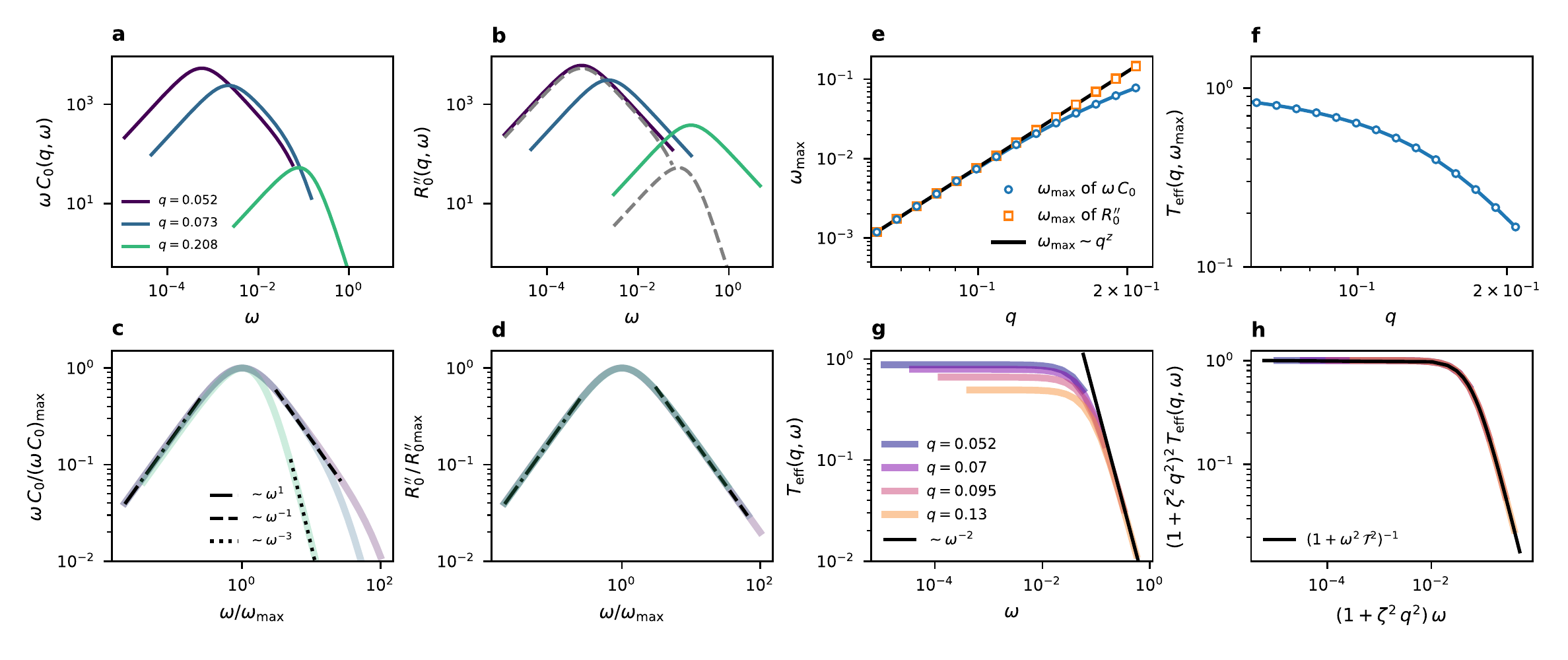}
\caption{\label{fig:wide3} \textbf{Relaxation spectra and effective temperature in the colored-noise driven field theory.} (\textbf{a}) Fluctuation spectra as a function of frequency at various $q$-s (colored curves, see legend). (\textbf{b}) Frequency-resolved response function at the same $q$-s of (a) (colored curves, same legend as in (a)). The correlation spectra for the highest and lowest wave-vectors in (a) are reproduced here as dashed lines to show that the deviation between response and correlation is stronger at high $q$.
(\textbf{c}) Scaled fluctuation spectra from (a) revealing a deviation from scaling in the high-$q$/high-$\omega$ regime. The asymptotic behaviors of these functions are highlighted by power-law fits (straight lines, see legend).
(\textbf{d}) Scaled response functions from (b). Differently from (c) a good collapse of the data is observed for the response and the asymptotic behaviors are evidenced by power law fits (straight lines, same legend as in (c)). 
(\textbf{e})
Peak frequency of the correlator (circles) and of the response (squares) as a function of $q$. The straight full line corresponds to the equation $\omega_{\rm max}=\gamma \, q^4$ with $\gamma\approx 10^2$ (see text).  
(\textbf{f})
Effective temperature evaluated at the correlator peak-frequency as a function of $q$. 
(\textbf{g})
Frequency-resolved effective temperature in the low-$q$ regime (colored curves), different colors indicate different $q$-values (see legend). The black line represents the power-law $\omega^{-2}$ predicted by theory for the high-$\omega$ decay of $T_\mathrm{eff}$.
(\textbf{h})
Data from (g) are scaled according to the theory. 
}
\end{figure*}

We now show that the results of the previous paragraph can be rationalized using an appropriate non-equilibrium relaxation dynamics of a conserved scalar field $\varphi(x,t)$ at position $x=(x_1,..,x_d)$ in the $d$-dimensional space at time $t$. 
The dynamics of the field is governed  by the equations:

\begin{align} \label{eq:ModelB_0}
    \partial_t \varphi(x,t) &= -\gamma (i \nabla)^2 \frac{\delta H_{LG}}{\delta \varphi(x,t)} + \eta(x,t)\\
    \langle \eta(x,t) \, \eta(y,t^\prime) \rangle &= (i \nabla)^2 K_{\zeta,\mathcal{T}}(|x-y|,|t-t^\prime|)
    \label{eq:ModelB_noise}
\end{align}

\noindent where the parameter $\gamma$ represents a macroscopic mobility (setting the time-scale of the relaxation dynamics) and $H_{LG}$ is the standard $\varphi^4$ Landau-Ginzburg Hamiltonian

\begin{align} \label{eq:LG}
    H_{LG}[\varphi] = \int d^dx \, \left[ \frac{1}{2} (\nabla \varphi)^2 + \frac{r}{2} \varphi^2 + \frac{u}{4!} \varphi^4\right] \;.
\end{align}

\noindent In Eq.~(\ref{eq:ModelB_noise}) the kernel $K_{\zeta,\mathcal{T}}(|x-y|,|t-s|)$ indicates that the noise $\eta(x,t)$ is time and space translational invariant and correlated over a length-scale $\zeta$ and over a time-scale $\mathcal{T}$.
If we further assume that the noise relaxes exponentially both in space and in time then, in the Fourier domain, the noise kernel takes the Lorentzian form 

\begin{align}\label{eq:ModelB}
K(\kappa,\kappa') \!=\! (2 \pi)^{d+1} \frac{2 \, \gamma \, T}{(1 + \zeta^2 q^2)^2 + \omega^2 \, \mathcal{T}^2}  \, \delta(\kappa + \kappa^\prime) \;    
\end{align}

\noindent where $\kappa$ compactly indicates~\cite{tauber2014critical} the Fourier-vector ${\kappa = (\mathbf{q},\omega)}$ and $T$ is the noise strength.
As usual we assume that $r\geq 0$ ($r$ vanishes at the Gaussian critical point) and $u>0$. 
These type of models have been studied in the past (but only in the case of a non-conserved order parameter~\cite{garcia1994colored}) as the simplest colored-noise driven field-theories. 
Note that by taking the limit $\mathcal{T}\rightarrow 0$ and $\zeta\rightarrow 0$ the model defined by  Eq.s~(\ref{eq:ModelB_0})-(\ref{eq:ModelB}) reduces to the standard version of Model B which describes the relaxational dynamics of a conserved scalar field in equilibrium at a temperature $T$~\cite{tauber2014critical}. More details about the field theoretical framework in presence of the correlated noise field are given in Appendix~\ref{intro_col_field}.
 
\subsection*{Gaussian theory}
We start by showing that, already at the Gaussian level ($u=0$), the model (\ref{eq:ModelB_0})-(\ref{eq:ModelB}) predicts a scenario in qualitative agreement with the simulation data. To show this, we solve Eq.s~(\ref{eq:ModelB_0})-(\ref{eq:ModelB}) and find correlator and response (details are given in Appendix~\ref{gauss_model_app}) as,

\begin{align} 
    \label{G0}
    G_0(q,\omega) &= 
    \frac{1}{\gamma  q^2 \left(q^2+r\right)
    -i \, \omega }
    \\
    \label{R0}
    R_0(q,\omega) &= 
    \gamma \, q^2 \, G_0(q,\omega)
    \\
    \label{C0}
    C_0(q,\omega) &= \frac{2 \, \gamma  \, q^2 \, T \, G_0(-q,-\omega)\,G_0(q,\omega)}{\left(1+\zeta ^2 q^2\right)^2+ \omega^2 \mathcal{T}^2} 
\end{align}

\noindent where $G_0(q,\omega)$ is the standard Gaussian response propagator and the subscript zero refers to the Gaussian theory.
To understand the behavior of response and correlation functions given by Eq.s~(\ref{R0}) and (\ref{C0}) and how it compares with the numerical results of the previous paragraph, we plot $\omega \, C_0(q,\omega)$ and $R''_0(q,\omega)$ in Figs.~\ref{fig:wide3}(a) and (b), respectively. Since we are interested in the critical point we set $r=0$. As we will see in the following the peak frequency of $R''_0(q,\omega)$ and $\omega\,C_0(q,\omega)$ for low $q$ is given by $\omega_\mathrm{max}=\gamma \, q^4$, therefore we choose $\gamma$ such that the $\omega_\mathrm{max}$ of the Gaussian colored field-theory matches the one found in particles simulations (this yields $\gamma\approx 10^2$). The noise parameters are  set to $\mathcal{T}=16.47$ and $\zeta=5.59$ (this particular choice will be justified in the following).

In Fig.~\ref{fig:wide3}(a),(b) we plot $\omega\,C_0(q,\omega)$ and $R''_0(q,\omega)$ at the same ${q,\omega}$-values of the numerical simulations. The theoretical results for $\omega\,C_0(q,\omega)$ and $R''_0(q,\omega)$ show a striking similarity to the numerical results of Figs.~\ref{fig:wide1}(a),(b), i.e. the correlator and the response are considerably different at large $q$ but they almost coincide at low $q$-values. As in simulations we observe some deviations between $\omega \, C_0(q,\omega)$ and $R''_0(q,\omega)$ also at low $q$ if $\omega$ is high enough (see dashed lines in Fig.~\ref{fig:wide3}(b)). Furthermore, the shape of the scaled $\omega \, C_0(q,\omega)$, shown in Fig.~\ref{fig:wide3}(c), displays an evolution that is remarkably similar to the numerical one of Fig.~\ref{fig:wide1}(c). The correlator shows a fast ($\sim \omega^{-3}$) decay at large $q$, while it shows a slower decay ($\sim \omega^{-1}$) at intermediate and low $q$ in full analogy with the numerical results of Fig.~\ref{fig:wide1}(c). Also, for the scaled theoretical response of Fig.~\ref{fig:wide3}(d), the agreement with the numerical results (Fig.~\ref{fig:wide1}(d)) is evident as a good data-collapse is obtained.

From Eq.~(\ref{R0}) we readily find that  $\omega_\mathrm{max}=\gamma \, q^4$ for the response function, as also shown in Fig.~\ref{fig:wide3}(e) (squares). Differently for $\omega \, C_0(q,\omega)$, the function $\omega_\mathrm{max}(q)$ is more complicated but it can be approximated at low $q$ as ${\omega_\mathrm{max} \approx \gamma \, q^4-2 \gamma ^3 \, \mathcal{T}^2 \, q^{12} }$.
This implies that relaxation frequency for the correlator is slightly lower than the one of the response as shown also in Fig.~\ref{fig:wide3}(e) (circles) at large $q$-values in analogy with Fig.~\ref{fig:wide2}(a). Note, however, that the field-theory cannot capture the deviation from the perfect scaling $\omega_\mathrm{max}\sim q^z$ for the response function observed both in numerical simulations and also in the equilibrium lattice gas (see Fig.~\ref{fig:wide2}(a) and its inset). Indeed the response (\ref{R0}), for $r=0$, represents a truly scale-free relaxation in which the microscopic details of the dynamics have been completely washed-out by the coarse-graining. 

By using (\ref{R0}) and (\ref{C0}) in the definition (\ref{eq:teffdef}) we find the effective temperature of the Gaussian model (see also Appendix~\ref{teff_app}):

\begin{align} \label{eq:Teff}
    T_\mathrm{eff}(q,\omega) &= \frac{\omega \, C_0(q,\omega)}{ 2\, T \, R''_0(q,\omega) } = \frac{1}{(1 + \zeta^2 \, q^2)^2 + \omega^2 \,  \mathcal{T}^2} \; .
\end{align}

\noindent We first notice that the effective temperature (\ref{eq:Teff}) contains the essential information about the noise kernel of Eq.~(\ref{eq:ModelB}). This implies that, according to this model, the study of $T_\mathrm{eff}$ allows us to characterize the noise-field. In analogy with the analysis of Fig.~\ref{fig:wide2}(b) we report $T_\mathrm{eff}$ evaluated at the $\omega_\mathrm{max}$ of the correlator in Fig.~\ref{fig:wide3}(f). We find that this has the same tendency, observed in simulations, to approach unity from below as $q$ decreases. In the regime where $\omega_\mathrm{max}=\gamma q^4$, Eq.~(\ref{eq:Teff}) implies that, at low $q$-values, we can approximate ${T_\mathrm{eff}(q,\omega_\mathrm{max})\approx(1 + \zeta^2 \, q^2)^{-2}}$. We use this formula to fit the low-$q$ data of Fig.~\ref{fig:wide2}(b) to estimate the parameter $\zeta$. We find that the characteristic length-scale of the noise field $\zeta=5.897(0.073)$. Interestingly this value is close to the characteristic length of the velocity correlation studied in~\cite{maggi2021universality} for the microscopic critical active system. This suggests that the coarse-grained noise field embodies the velocity correlation that develops at the particle level.
We report the $T_\mathrm{eff}$ dependence on $\omega$ at different $q$-s  in Fig.~\ref{fig:wide3}(g).
Similarly to the numerical results in Fig.~\ref{fig:wide2}(c), at low  $\omega$, $T_\mathrm{eff}$ is approximately constant (its value approaches unity upon lowering $q$) while  $T_\mathrm{eff}$ decays as $\omega^{-2}$ at high $\omega$. Given the strong analogies between the theory and the numerical model, we use Eq.~(\ref{eq:Teff}) to fit the data Fig.~\ref{fig:wide2}(c). In this way we estimate $\mathcal{T}=16.47(0.25)$ and $\zeta=5.592(0.028)$ (this justifies the choice of parameters mentioned at the beginning of this paragraph). Although we do not attempt here a microscopic derivation of the parameter $\mathcal{T}$, we note that this is very close to the relaxation time of the microscopic active force. Finally we note that Eq.~(\ref{eq:Teff}) also implies that $T_\mathrm{eff}$ data for different $q$ and $\omega$ should collapse on the same curve $(1+\mathcal{T}^2\omega^2)^{-1}$ when we plot $(1+\zeta^2 \, q^2)^2 \, T_\mathrm{eff}(q,\omega)$ as a function of $(1+\zeta^2 \, q^2)\,\omega$  (as shown in Fig.~\ref{fig:wide3}(h)). Applying the same procedure to the data in Fig.~\ref{fig:wide2}(c) we find a good data collapse as shown in Fig.~\ref{fig:wide2}(d).

At this point, two questions naturally arise: (\textit{i}) since the Gaussian theory is strictly valid only for $d$ above the critical dimension $d=d_c$ (which turns out to be $d_c=4$ as in equilibrium) why does the $T_{\mathrm{eff}}$ of Eq.~(\ref{eq:Teff}) fit well the simulation data in $d=2\,$? (\textit{ii}) is the colored-noise driven field theory consistent with the  critical exponent of the Ising universality class in $d<4\,$?
To  partially answer these questions we then study the field theory perturbatively in $d=4-\epsilon$ dimensions as detailed in the next paragraph.

\subsection*{Results from perturbation theory}

To show that the $T_\mathrm{eff}(q,\omega)$ given by Eq.~(\ref{eq:Teff}) also applies below $d_c$, we calculate the correlator and the response to first order in perturbation theory (see Appendix~\ref{teff_app}). By inserting these functions in the definition (\ref{eq:teffdef}), we find the same exact result of Eq.~(\ref{eq:Teff}). The key steps leading to this conclusion can be summarized in the following way. We start by writing the Dyson equation of the form $G(q,\omega) = \left[G_0(q,\omega)^{-1} + \Sigma(q,\omega)\right]^{-1}$ with $\Sigma(q,\omega)$ being the self-energy \cite{ZinnJustin}. Then the perturbation theory to one loop yields 

\begin{equation}
\label{R1loop}
R(q,\omega) = R_0(q,\omega ) \left[1-\frac{u}{2} \, R_0(q,\omega )
   \, \mathcal{I}(\zeta ,\mathcal{T},r)\right]
\end{equation}

\noindent where the analytical expression of the integral $\mathcal{I}$ is provided in Appendix~\ref{teff_app}. 
The corresponding correlation function reads
\begin{equation}
 C(q,\omega) = 
C_0(q,\omega ) \left[1- u \, R'_0(q,\omega ) \; \mathcal{I}(\zeta
   ,\mathcal{T},r)\right]
\end{equation}

\noindent where the prime indicates the real part of $R_0$ 
The crucial observation here is that the response (\ref{R1loop}) depends on the correlated-noise parameters only via the term $\mathcal{I}$,
while the Lorentzian kernel of the noise
\textit{factors out} in $C(q,\omega)$ as in $C_0(q,\omega)$.
Therefore, $T_\mathrm{eff}(q,\omega)$ turns out to be the same as the one computed in the Gaussian theory. This suggests that the expression (\ref{eq:Teff}) can be used as a suitable approximation to model the effective temperature also below the upper critical dimension.

We now turn to the question regarding the universality of the colored-noise driven field theory. A scaling analysis of the model readily shows that, under the Kadanoff transformation $x \to b x$ and $t \to b^z t$, we have for the couplings $r^\prime = r \, b^2, \; u^\prime = u \, b^{4-d}, \; \zeta^\prime = \zeta \, b^{-1}$, and $\mathcal{T}^\prime = \mathcal{T} \, b^{-z}$ with $z=4$ (see Appendix~\ref{scal_app}). While it is evident from the scaling analysis that $d_c=4$, it is also clear that $\mathcal{T}$ and $\zeta$ are irrelevant operators in the RG sense at the Gaussian fixed-point. This also suggests that the non-Gaussian fixed point is the usual Wilson-Fisher fixed point. In order to check this, we employ Wilson's RG up to one loop (further details are provided in Appendix~\ref{RG_app}).  
With $\epsilon=4-d$, the calculation, that does not require any small $\mathcal{T}$ and $\zeta$ approximation, brings to the non-Gaussian fixed point ${r^\prime_{F.P.}=-\epsilon \,  \mathcal{C}(\Lambda,\zeta^\prime,\mathcal{T}^\prime) / 6}$ and ${u^\prime_{F.P.}= \epsilon \, 16 \pi^2 \, B(\Lambda,\zeta^\prime,\mathcal{T}^\prime)^{-1} / 3}$. On the critical surface  $\zeta^\prime_{F.P.}=\mathcal{T}^\prime_{F.P.}=0$ and we obtain $\mathcal{C}(\Lambda,0,0)=\Lambda^2$, $B(\Lambda,0,0)=1$, i. e., the Wilson-Fisher fixed point \cite{le1991quantum}. 


This study confirms the idea that the universality class of the colored noise driven field theory is the same as the Ising universality class which is compatible with the observation reported in the present work and in previous investigations~\cite{PhysRevLett.123.068002,maggi2021universality}.

\section*{Conclusions}

In this work we have studied numerically and analytically the dynamical properties of an active system around its MIPS critical point. 
We have found that the FDT is strongly violated at short time and length scales and that effective equilibrium is progressively restored at large spatiotemporal scales.

It is interesting to qualitatively compare this scenario with the breakdown of the FDT found in other types of non-equilibrium systems. For example in glassy systems, the FDT violation is typically stronger at large spatiotemporal scales~\cite{BerthierPRL98}, which can be interpreted as the rapid equilibration of the fast degrees of freedom at the bath temperature followed by a gradual re-equilibration of the slow degrees of freedom (associated with the collective rearrangements) which ``remember'' another temperature. Contrarily, in our critical active system the stronger violation occurs at small length- and time-scales highlighting a very different type of non-equilibrium behavior.

We have found that the the frequency dependence of the correlator and the response in the particle simulations cannot be rationalized in terms of a field theory driven by white noise.
Differently we have shown that the scenario resulting from simulations is captured by a field-theory where the order-parameter dynamics is driven by a noise field correlated in time and space. The model  directly derives from the numerical observation that the spontaneous fluctuations appear to be weaker (at high $q$ and $\omega$) than the ones induced by the external field. In this context the correlated noise field turns out to be a crucial ingredient to model the critical dynamics at the coarse-grained level, as the theory ---already in its Gaussian version--- qualitatively reproduces most of the non-equilibrium features observed numerically. 
In particular the Gaussian model predicts a scale-dependent effective temperature $T_\mathrm{eff}(q,\omega)$ that tends to a constant for ${(q\rightarrow 0, \omega\rightarrow 0)}$ and therefore equilibrium is restored asymptotically.
We remark that this is consistent
with our previous numerical results~\cite{maggi2021universality} showing that the Ising critical exponents are observed for large system sizes. 
In addition our model also phenomenologically justifies the deviations from the Ising exponents found in~\cite{siebert2018critical} since the spatial correlations of the noise are important on a small scale and may affect the static critical exponents. 

Finally, by using a dynamic RG approach, we have found that the colored-noise driven field-theory falls in the Ising universality class also below the upper critical dimension. This happens because the noise memory kernel introduces two operators that are RG irrelevant. It is worth noting that, in the present work, the analytical computation has been done using Wilson's RG scheme up to one loop. A detailed study of our field theoretical framework, at higher orders in perturbation theory, is in progress also to understand how the effective temperature may change its functional form beyond first order. As a further perspective it would be interesting to measure correlation and response functions also in the critical active lattice models~\cite{PhysRevLett.123.068002,dittrich2021critical} and to check if the current scenario applies also in these cases. Although no field-free method has been yet developed for measuring the response in these systems, it could be that their numerical efficiency still allows one to obtain quickly the response function by directly applying the field. Moreover, it would be also interesting to explore the consequences of having a correlated noise field deep in the phase separation region of the MIPS and to understand, for example, what the role such a complex noise may have in the formation of interfaces at a coarse-grained level.
Finally, it would be worth trying to derive the correlated-noise driven field-theory directly from the microscopic dynamics by using, for example, the Mori-Zwanzig formalism~\cite{zwanzig2001nonequilibrium} or by applying the Ito's lemma to the density field as done in~\cite{dean1996langevin} for equilibrium systems.

\section*{Acknowledgments.}
M.P. has received funding from the European Union's Horizon 2020 research and innovation programme under the MSCA grant agreement No 801370
and by the Secretary of Universities 
and Research of the Government of Catalonia through Beatriu de Pin\'os program Grant No. BP 00088 (2018).






\setcounter{equation}{0}
\setcounter{figure}{0}


\bibliography{mpbib}

\clearpage

\appendix

\section*{Appendix}

\subsection{Implementation of the Malliavin weights for AOUPs.} \label{MalliApp}

To evaluate the response over unperturbed trajectories we follow Ref.~\cite{szamel2017evaluating} according to which the response function of interest is given by


\begin{equation}
\label{malli}
[\partial_h \langle \rho_{\mathbf q}(t) \rangle]_{h\rightarrow 0} = \langle \rho_{\mathbf q}(t) (Q(t)+P(t))\rangle \\
+ \tau \, \langle \dot{\rho}_{\mathbf q}(t) Q(t)\rangle
\end{equation}

\noindent where the Malliavin variables $Q$ and $P$ can be rewritten in terms of single-particle variables, i.e. $Q=\sum_i Q_i$ and $P=\sum_i P_i$. The evolution of the $Q_i$ and $P_i$ is governed by the equations:

\begin{eqnarray}
\label{Qeq}
    \dot{Q}_{i} &=&-\frac{\mu^3}{D}\sin(\mathbf{q}\cdot\mathbf{r}_i(t))\, \mathbf{q}\cdot\boldsymbol{\xi}_i\\
    \label{Peq}
    \dot{P}_{i} &=&\frac{\tau\mu^3}{2D}\dot{\mathbf{x}}_i\cdot\mathbf{H}_i \cdot \boldsymbol{\xi}_i
\end{eqnarray}

\noindent where $\mathbf{H}_i$ is the Hessian matrix $\mathbf{H}_i= -2 \, \mathbf{q} \otimes \mathbf{q} \, \cos(\mathbf{q}\cdot \mathbf{r}_i(t))$ defined by the dyadic product $\mathbf{q} \otimes \mathbf{q}$. To compute the response (\ref{malli}), Eq.s (\ref{Qeq}) and (\ref{Peq}) must be integrated according to the Ito's rule from $t\geq0$ with initial conditions $Q(t\leq0)=0$ and $P(t\leq0)=0$. This method is particularly convenient first because the white noise $\boldsymbol{\xi}_i$ in Eq.s (\ref{Qeq}) and (\ref{Peq}) is the same as the one in the particle's dynamics (Eq.s~(\ref{micro1}) and (\ref{micro2})) and no additional random numbers have to be generated. Secondly Eq.s (\ref{Qeq}) and (\ref{Peq}) can be efficiently evaluated in parallel for all particles and summed to compute Eq.~(\ref{malli}) only when needed.

\subsection{Critical dynamics with correlated noise}
\label{intro_col_field}

In order to consider both cases of non conserved ad conserved scalar order parameter, we start
with the following relaxation non-equilibrium dynamics 
\begin{align} \label{eq:modelAB}
    \partial_t \varphi(x,t) &= -\gamma (i \nabla)^a \frac{\delta H_{LG}}{\delta \varphi(x,t)} + ( -\nabla )^{a/2}\eta(x,t) \nonumber \\ 
    H_{LG}[\varphi]         &= \int d^dx \, \left[ \frac{1}{2} (\nabla \varphi)^2 + \frac{r}{2} \nonumber \varphi^2 + \frac{u}{4 !} \varphi^4  \right] \\ 
    \langle \eta(x,t) \eta(y,t^\prime) \rangle &=  (i \nabla)^a K_{\zeta,\mathcal{T}}(|x-y|,|t-t^\prime|)
\end{align}
with $a=0,2$ for Model A and B, respectively.
The noise field $\eta(x,t)$ is a {\it fast degree of freedom} but still correlated over a (non-zero) length-scale $\zeta$ a time-scale $\mathcal{T}$. 
Its dynamics can be represented through the following  Ornstein-Uhlenbeck evolution~\cite{garcia1994colored} 
\begin{equation}
    \mathcal{T} \, \dot{\eta}(x,t) =  - (i \nabla)^a (1 - \zeta^2 \nabla^2) \eta(x,t) + \xi(x,t)
\end{equation}
with
\begin{align}
\langle \xi(x,t) \rangle &= 0\; , \\ \nonumber
\langle \xi(x,t) \xi(y,t^\prime)\rangle &= 2 \gamma T \delta^{(d)}(x-y) \delta(t-t^\prime) \; .
\end{align}

After standard manipulations \cite{PhysRevA.8.423,dominicis1976technics,janssen1976lagrangean,PhysRevB.18.353,jensen1981functional,tauber2014critical}, we arrive to the Janssen–De Dominicis 
dynamical action $\mathcal{A}[\hat{\varphi},\varphi] = \mathcal{A}_0 + \mathcal{A}_{int}$ that is
\begin{widetext}
\begin{align} \nonumber
    \mathcal{A}_0[\hat{\varphi},\varphi]     &= -\frac{1}{4} \int d^dx \,d^dy \, dt \, dt^\prime \, \hat{\varphi}(x,t) \left[ (i \nabla)^a K_{\zeta,\mathcal{T}}(|x-y|,|t-t^\prime|)\right] \hat{\varphi}(y,t^\prime) + \int d^dx\, dt\, \hat{\varphi}(x,t) \left[ \partial_t + \gamma (i\nabla)^a (r - \nabla^2) \right] \varphi(x,t)\\
    \mathcal{A}_{int}[\hat{\varphi},\varphi] &= \frac{\gamma u}{3 !} \int d^dx\,dt\, \hat{\varphi}(x,t) (i \nabla)^a \varphi(x,t)^3 \; .
\end{align}
\end{widetext}
with $\hat{\varphi}(x,t)$ indicating the response field.
We introduce the Fourier transform of the fields
\begin{align}
\varphi(q,\omega) = \frac{1}{(2 \pi)^{d+1}} \int d^dx\,dt\, e^{-i (qx - \omega t)} \varphi(x,t)  \\ \nonumber
\hat{\varphi} (q,\omega) = \frac{1}{(2 \pi)^{d+1}} \int d^dx\,dt\, e^{-i (qx - \omega t)} \hat{\varphi}(x,t)
\end{align}
The dynamical action in the Fourier space is
\begin{widetext}
\begin{align}
    \mathcal{A}_{0}[\hat{\varphi},\varphi]   &= -\frac{\gamma}{2} \int d\kappa \hat{\varphi}(-\kappa)  q^{a} M(\kappa)\hat{\varphi}(\kappa)   + \int d\kappa \,\hat{\varphi}(\kappa) \left[ -i \omega + \gamma q^a (q^2 + r)\right] \varphi(\kappa) \\ \nonumber 
\mathcal{A}_{int}[\hat{\varphi},\varphi] &= \frac{\gamma u}{3} \int d\kappa_1 \,d\kappa_2 \,d\kappa_3 \, \hat{\varphi}(\kappa_1) \varphi(\kappa_2) \varphi(\kappa_3) \varphi(-\kappa_1 - \kappa_2 - \kappa_3) q_1^{a} \\ \nonumber
M(\kappa) &\equiv \; \frac{T}{(1 + \zeta^2 k^2)^2 + \omega^2 \mathcal{T}^2} \; ,
\end{align}
\end{widetext}
where we are adopting the notation
 $\kappa = (q,\omega)$ with $q$ the wave vector 
 ($q \in \mathbb{R}^d$) and $\omega \in \mathbb{R}$
the frequency, $\int d\kappa \equiv \int_{0}^{\Lambda} k^{d-1} \frac{dk}{(2 \pi)^d} \int_{-\infty}^{+\infty}\frac{d\omega}{2 \pi}$ with $\Lambda = \frac{2 \pi}{a}$ the ultraviolet cutoff. In the following we consider the behavior of the system close to the critical point, we assume that $r\geq0$ and vanishes at the critical point and we set the strength of the noise to $T=1$.


In order to compute the response function we couple consider also the case in which the $\varphi$ is coupled to an external field $h$ and thus the Hamiltonian becomes
\begin{align}
    H_h[\varphi] &= H_{LG}[\varphi] + \int d^dx \, \varphi \, h \; 
\end{align}
The response function $R(x-x^\prime,t-t^\prime)$ is defined as follows \cite{tauber2014critical}
\begin{align}
    R(x-x^\prime,t-t^\prime) &= \left. \frac{\delta \langle \varphi(x,t) \rangle}{\delta h(x^\prime,t^\prime)} \right|_{h=0}
    = \gamma \langle \varphi(x,t) (i \nabla )^a \hat{\varphi}(x^\prime, t^\prime) \rangle\; .
\end{align}
In Fourier space, because of the translational and rotational invariance, we get
\begin{align}
    R(q,\omega) &= \gamma q^a G(q,\omega)
\end{align}
with $G(q,\omega)$ the response propagator of the theory.

\subsection{Gaussian Model}
\label{gauss_model_app}

Neglecting the non-linear interaction we obtain the Gaussian model whose generating functional can be written as
\begin{align*}
    Z_{dyn}[\boldsymbol{J}] &= \int \mathcal{D}[\boldsymbol{\Phi}] \, e^{-\frac{1}{2} \int d\kappa \, \boldsymbol{\Phi}^\dagger (-\kappa) \mathbf{G_0}(\kappa) \boldsymbol{\Phi} \kappa + \int d\kappa \, \boldsymbol{J}(\kappa) \cdot \boldsymbol{\Phi} \kappa} \\ 
    \; ,
    \mathbf{G_0}^{-1}(\kappa) &=     
    \begin{bmatrix}
     -2 \gamma q^a M(\kappa)&  -i \omega + \gamma q^a (q^2 + r)\\
     i \omega + \gamma q^a (q^2 + r) & 0 
    \end{bmatrix}
\end{align*}
where we have introduced a doublet of scalar field ${\boldsymbol{\Phi}\equiv (\hat{\varphi}(\kappa),\varphi)(\kappa) )}$ and a doublet of external sources ${\boldsymbol{J}\equiv(\hat{J}(\kappa),J(\kappa))}$. 
The matrix $\mathbf{G}_0$ contains the inverse of the free propagator of the theory.
Performing the Gaussian integral we get
\begin{align}
    Z_{dyn}[\boldsymbol{J}] &= Const. \times e^{\frac{1}{2} \int d\kappa \, \boldsymbol{J}^\dagger(-\kappa) \boldsymbol{G_0} (\kappa) \boldsymbol{J}(\kappa)} \; .\end{align}
The free propagators can be computed by solving the set of equations $\boldsymbol{G_0}^{-1} \boldsymbol{G_0} = \boldsymbol{1} $ so that
\begin{align}
    \boldsymbol{G_0}^{-1} &=
        \begin{bmatrix}
     0 &  \frac{1}{i \omega + \gamma q^a (q^2 + r)}\\
     \frac{1}{-i \omega + \gamma q^a (q^2 + r)} & \frac{2 \gamma q^a M(\kappa)}{\omega^2 + \left[ \gamma q^a (q^2 + r)\right]^2}  
    \end{bmatrix} \; .
\end{align}
The two-point correlation functions of the Gaussian model are
\begin{align*}
    \begin{bmatrix}
     \langle \hat{\varphi}(\kappa) \hat{\varphi}(\kappa^\prime)\rangle  &  \langle \hat{\varphi}(\kappa)\varphi(\kappa^\prime)\rangle \\
     \langle \varphi(\kappa)\hat{\varphi}(\kappa^\prime)\rangle & \langle \varphi(\kappa)\varphi(\kappa^\prime)\rangle  
    \end{bmatrix} &= (2 \pi)^{d+1}\boldsymbol{G_0} (\kappa) \delta(\kappa + \kappa^\prime) \; .
    \end{align*}
The Gaussian theory leads thus to the following expression for the free propagators of the theory, i. e., the response propagator $G_0$ and the correlation function $C_0$, that are
\begin{align}
    G_0(q,\omega) &= \frac{1}{\gamma q^a (q^2 + r) - i \omega}\\ \nonumber
    C_0(q,\omega) &= \gamma q^a G_0(-q,-\omega) M(q,\omega) G_0(q,\omega) \; .
\end{align}
The response function of the Gaussian theory is $R_0(q,\omega) = \gamma q^a G_0(q,\omega)$.

\subsection{Dyson's equation}
\label{teff_app}

We can rewrite correlation and response function of the free theory in the following way
\begin{align}
    G_0(q.\omega) &= \frac{\Delta + i \omega}{\Delta^2 + \omega^2} \\
    C_0(q,\omega) &= \frac{\gamma 2 q^a M(q,\omega)}{\Delta^2 + \omega^2} = \frac{2 \gamma q^a}{\omega}  G_0^{\prime\prime}(q,\omega) M(q,\omega) \\
    \Delta &\equiv \gamma q^a (q^2 + r) \; .
\end{align}
We immediately realize that the memory kernel $K(q,\omega)$ plays the role of an effective scale-dependent effective temperature
\begin{align}
     M(q,\omega) &= \frac{\omega \, C_0(q,\omega)}{ G_0^{\prime\prime}(q,\omega) 2 \gamma q^a} \; .
\end{align}
For a generic theory, the response function $R(q,\omega)$ can be computed using the Dyson equation
\begin{align}
    G(q,\omega) = \frac{1}{G^{-1}_0 + \Sigma}
\end{align}
with $\Sigma$ being the self-energy \cite{tauber2014critical,ZinnJustin}. Writing ${\Sigma = \Sigma^\prime + i \Sigma^{\prime\prime}}$, we get
\begin{align}
    G(q,\omega) &= \frac{(\Delta + \Sigma^\prime) + i (\omega - \Sigma^{\prime\prime})}{(\Delta + \Sigma^\prime)^2 + (\omega - \Sigma^{\prime\prime})^2} \\
    C(q,\omega) &= \frac{2\gamma q^a M(q,\omega)}{(\Delta + \Sigma^\prime)^2 + (\omega - \Sigma^{\prime\prime})^2}
\end{align}
Combining these expressions we obtain
\begin{align}
     M(q,\omega) &= \frac{(\omega - \Sigma^{\prime\prime}) \, C(q,\omega)}{ G^{\prime\prime}(q,\omega)} 
\end{align}
By employing perturbation theory, the one-loop computation of the self-energy brings to
\begin{widetext}
\begin{align}
    G(q,\omega) &= G_0(q,\omega) \left[ 1 - \frac{u}{2} \gamma q^a G_0(q,\omega) \, \mathcal{I}(\zeta,\mathcal{T},r)\right] + O(u^2) \\
    \mathcal{I}(\zeta,\mathcal{T},r) &\equiv \int_{0<|k|<\Lambda} \frac{d^d k}{(2 \pi)^d} \frac{k^a}{\left(\zeta^2  k^2+1\right) k^a\left(k^2+r\right) \left(\gamma  \mathcal{T}  k^a\left(k^2+r\right)+\zeta^2  k^2+1\right)} \; .
\end{align}
\end{widetext}
Since $C(q,\omega)=G(q,\omega)G(-q,-\omega)M(q,\omega)$, up to first order in $u$ we get 
\begin{align}
    C(q,\omega) = C_0(q,\omega) \left[ 1 - u \, \gamma q^a \, G_0^\prime \, \mathcal{I}(\zeta,\mathcal{T},r) \right] + O(u^2) \; .
\end{align}
From the Dyson equation it follows that $\Sigma=\frac{u}{2} \mathcal{I}(\zeta,\mathcal{T},r)$,
thus we have $\Sigma^{\prime\prime}=0$ 
so that
\begin{align}
     M(q,\omega) &= \frac{\omega \, C(q,\omega)}{ G^{\prime\prime}(q,\omega) 2 \gamma q^a} \; .
\end{align}

\subsection{Scaling Analysis}
\label{scal_app}

We perform the following scaling transformations in the real space
\begin{align}
    x &\to b \, x \\ \nonumber
    t & \to b^{z} \\ \nonumber
    \varphi(x,t) & \to b^\chi \varphi \\ \nonumber
    \hat{\varphi}(x,t) & \to b^{\hat{\chi}} \varphi 
\end{align}
that in Fourier space become
\begin{align}
    q &\to b^{-1} q \\ \nonumber
    \omega &\to b^{-z} \omega \\ \nonumber
    \varphi(k,\omega) & \to b^{\chi + d + z} \varphi(k,\omega) \\ \nonumber
    \hat{\varphi}(k,\omega) & \to b^{\hat{\chi} + d + z} \varphi(k,\omega) \; .
\end{align}

From the scaling analysis we get
\begin{align}
    \gamma^\prime &= \gamma \, b^{d+z+2 \tilde{\chi} - a} \\ \nonumber
    \mathcal{T}^\prime &= \mathcal{T}\,  b^{-z} \\ \nonumber
    \zeta^\prime &= \zeta \, b^{-1} \\ \nonumber
    u^\prime &= u \, b^{2 z - d - 2 a} \\ \nonumber
    r^\prime &= r \, b^{z-a} \; .
\end{align}
As usual we have $\tilde{\chi}+\chi+d=0$ and thus $z=2+a$ meaning that $z=2$ for model A, and $z=4$ in model B. The couplings $r$ and $u$ scale in the same way in both models
\begin{align}
    r^\prime &= r \, b^2 \\ \nonumber
    u^\prime &= u \, b^{4-d} \; .
\end{align}
Above $d=4$ the only relevant parameter is $r$. Around and below $d=4$, the coupling of the non-linear interaction becomes relevant.

\subsection{Non-Gaussian fixed point}
\label{RG_app}

In order to provide a quantitative analysis of the stability of the equilibrium non-Gaussian fixed point with respect to fluctuations due to correlated noise, we employ the usual RG machinery. 
This program can be accomplished using recursion relations near $d=4$ as in the case of static critical phenomena. 

As usual, we consider a scale parameter $b>1$ and the corresponding scale transformation $R_b^s$ that is a change of scale of factor $b$. The second operation $R_b^i$ that has to be applied to the diagrammatic expansions is the integration over internal wave vectors $k$ in the domain $b^{-1} \Lambda < |k| < \Lambda$ and internal frequency $\omega$ ranging from $-\infty < \omega < \infty$. The two commutative operations define the RG transformation $R = R_b^s R_b^i$. 
The renormalized coupling constants $r_R$ and $u_R$ we are looking for are then related with vertex functions $\Gamma^{1,1}(q,\omega)$ and $\Gamma^{1,3}(q,\omega)$ in the following way \cite{tauber2014critical}
\begin{align}
    r_R &= \lim_{q\to 0} \frac{1}{\gamma q^2} R_b^i \left[ \Gamma^{(1,1)}(q,0) \right] \\
    u_R &= \lim_{q \to 0} \frac{1}{\gamma q^2} R_b^i \left[\Gamma^{(1,3)}(q,0) \right] \; ,
\end{align}
and the corresponding RG equations shall be obtained by performing the scaling $r_R \to r^\prime$ and $u_R \to u^\prime$.  For computing the vertex functions $\Gamma(q,\omega)^{(a,b)}$ we employ diagrammatic perturbation theory up to one loop. The corresponding expressions of $\Gamma^{(1,1)}(q,0)$ and $\Gamma^{(1,3)}(q,0)$
are
\begin{align}
    \Gamma^{(1,1)}(q,0) &= \gamma q^{2} \left[ r + q^2 + \frac{u}{2} \mathfrak{I}_{B}(r)\right] \\
        \Gamma^{(1,3)}(q,0) &= \gamma  \left(\frac{3q}{2}\right)^2 u \left[ 1 - \frac{3}{2} u \, \mathfrak{J}_{B}(r,0)\right]
\end{align}
where we have defined the following quantities
\begin{widetext}
\begin{align}
    \mathfrak{I}_{B}(r) &= 2 \gamma \int_{\Lambda/b < |k| < \Lambda} \frac{d^dk}{(2 \pi)^d} \int_{-\infty}^{+\infty} \frac{d\omega}{2 \pi} \, \frac{1}{\Delta(k)^2 + \omega^2} \frac{k^2}{(1+\zeta^2 k^2)^2 + \omega^2 \mathcal{T}^2} = \\ \nonumber
    &= 2 \gamma \int_{\Lambda/b < |k| < \Lambda} \frac{d^dk}{(2 \pi)^d} \frac{1}{\gamma (r + k^2) (1 + \zeta^2 k^2) (1 + k^2 \zeta^2 + \Delta(k) \mathcal{T})}\\
    \mathfrak{J}_{B}(r,0) &= \int_{\Lambda/b < |k| < \Lambda} \frac{d^dk}{(2 \pi)^d} \frac{\gamma k^2 \left[ 1 + \zeta^2 k^2 + 2 \Delta(k) \mathcal{T} \right]}{2 \Delta(k)^2 (1 + \zeta^2 k^2) (1 + \zeta^2 k^2 + \Delta(k) \mathcal{T})^2}  \\
    \Delta(k) &\equiv \gamma k^2 (r + k^2) \; .
\end{align}
\end{widetext}
We can write the following RG equations for the renormalized couplings
\begin{align} \label{eq:RG1}
    r^\prime &= b^2 r_R \\ \label{eq:RG2} 
    u^\prime &= b^{4-d} u_R \; .
\end{align}
In order to write down (\ref{eq:RG1}) and (\ref{eq:RG2}) as differential equations, we introduce $s=\log b$, $\epsilon=4-d$,
and we finally get
\begin{align}
    \frac{d r^\prime}{d s} &= 2 r^\prime + \frac{u^\prime}{16 \pi^2} A(\Lambda,\zeta,\mathcal{T})\\
    \frac{d u^\prime}{d s} &= \epsilon u^\prime - \frac{3}{16 \pi^2} (u^\prime)^2 B(\Lambda,\zeta,\mathcal{T}) \\
\frac{d \zeta^\prime}{d s} &= - \zeta^\prime  \\ 
\frac{d \mathcal{T}^\prime }{d s} &= -4 \mathcal{T}^\prime \; , 
\end{align}
where we have defined the following functions
\begin{align}
    A(\Lambda,\zeta,\mathcal{T}) &\equiv \frac{\Lambda^2}{(1 + {\zeta^\prime}^2 \Lambda^2) (1 + {\zeta^\prime}^2 \Lambda^2 + \gamma^\prime \mathcal{T}^\prime \Lambda^4)} \\
B(\Lambda,\zeta,\mathcal{T}) &\equiv     \frac{1 + {\zeta^\prime}^2 \Lambda^2 + 2 \gamma^\prime \Lambda^4 \mathcal{T}^\prime}{(1 + {\zeta^\prime}^2 \Lambda^2) (1 + {\zeta^\prime}^2 \Lambda^2 + \gamma^\prime \mathcal{T}^\prime \Lambda^4)^2}
\; .
\end{align}
The fixed point equations for $r^\prime_{F.P.}$ and $u^\prime_{F.P.}$ are
\begin{align}
    r^\prime_{F.P.} &= -\frac{1}{6} \, \epsilon\, \mathcal{C}(\Lambda,\zeta^\prime,\mathcal{T}^\prime) \\
    u^\prime_{F.P.} &= \epsilon \, \frac{16 \pi^2}{3} B(\Lambda,\zeta^\prime,\mathcal{T}^\prime)^{-1} \\ 
    \mathcal{C} (\Lambda,\zeta^\prime,\mathcal{T}^\prime) &\equiv A / B \; .
\end{align}
Because under RG transformations $\zeta^\prime$ and $\mathcal{T}^\prime$ go to zero, we obtain $\mathcal{C}(\Lambda,0,0)=\Lambda^2 $ and $B(\Lambda,0,0)=1$
so that the non-Gaussian fixed point is the usual Wilson-Fisher fixed point.


\end{document}